\documentstyle[12pt,epsfig]{article}
\def\lapprox{\lower .7ex\hbox{$\;\stackrel{\textstyle <}{\sim}\;$}}

\def\GeV{{\rm GeV}}
\def\TeV{{\rm TeV}}
\def\ap{{a'}}
\def\vp{{v'}}

\def\shat{\hat{s}}
\def\qq{q \bar q}
\def\uu{u \bar u}
\def\dd{d \bar d}
\def\tt{t \bar t}
\def\pp{p \bar p}

\begin{document}
\begin{titlepage}
\vspace*{-1cm}
\begin{flushright}
DTP/96/24   \\
March 1996 \\
\end{flushright}                                
\vskip 2.cm
\begin{center}                                                                  
{\Large\bf The Effects of a New Vector Boson on the Top Quark Cross Section 
at the Tevatron}
\vskip 1.cm
{\large  T.~Gehrmann$^{a}$ and
W.J.~Stirling$^{a,b}$} 
\vskip .4cm
$^a$\ {\it Department of Physics, University of Durham \\
Durham DH1 3LE, England }\\
\vskip .2cm
$^b$\ {\it Department of Mathematical Sciences, University of Durham \\
Durham DH1 3LE, England }\\
\vskip 1cm                                                                    
\end{center}

\begin{abstract}
It has recently been shown that a new neutral vector boson $Z'$, with
a mass of order 1~TeV and weak couplings to quarks only, could explain
both the anomalous values of $R_b$ and $R_c$ and the apparent
excess of large $E_T$ jet events measured by the CDF collaboration.
We calculate the effects of  $Z'$ exchange on the $t\bar{t}$
production cross section at Tevatron $p \bar p$ collider
 energies, including next-to-leading-order
QCD corrections. We find a significant enhancement of the cross
section  and study the resulting $t \bar t$ invariant mass
distribution, which could provide a decisive test of 
the $Z'$ model.
\end{abstract}

\vskip 3cm
PACS: 14.70.Pw; 14.65.Ha; 12.38.Bx; 12.38.Qk.

\vskip 0.3cm
Keywords: New vector bosons; top quark physics; hadron--hadron collisions.
\vfill
\end{titlepage}                                                                
\newpage                                                                       

The predictions of the Standard Model are in impressive agreement
with a wide range of experimental measurements. There are, however,
several cases where  theory and data appear to be in disagreement:
the values of $R_b$ and $R_c$ measured at LEP are significantly
different from the Standard Model prediction \cite{rbrc}, and there
appears to be an excess of large $E_T$ jets in the CDF data
\cite{cdfjet}. It has recently been pointed out \cite{abfgm,clrv} that
both these effects could be explained by introducing a new U(1) gauge
boson ($Z'$) of mass $O(1~\mbox{TeV})$ which mixes at the $10^{-3}$ level
 with the $Z^0$ and has similar couplings to quarks.
By assuming universal couplings to the three fermion generations, and by
adjusting the couplings to the $u$--, $d-$quarks  and charged leptons,
the predictions for $R_b$ and $R_c$ can be brought into line
with experiment. The additional contributions to quark--(anti--)quark
scattering mediated by $s-$ and $t-$channel $Z'$ exchange increase
the predicted large $E_T$ jet rate at the Tevatron $p \bar p$ collider,
thus `explaining' the recent CDF data \cite{cdfjet}.

An important feature of the $Z'$ models of Refs.~\cite{abfgm,clrv} is that
$Z'$ vector and axial couplings to $u-$type quarks turn out to be quite
large. In fact the  effective $Z' u \bar u$ coupling is of the same
order as the strong coupling:
$ (\vp_u^2 + \ap_u^2) \alpha_W \sim O(10)\; \alpha_W \sim \alpha_S$,
which explains why the $Z'$ contribution to the large $E_T$ jet cross
section is comparable to  the QCD contribution.  Another
implication of this, which is the subject of the present study, is that
the top quark production cross section at the Tevatron collider
($\sigma_t$) is similarly enhanced, i.e.
the model gives rise to an additional `anomalous' contribution $\sigma'_t$
from $\qq\to (Z')^* \to \tt$ which is the same order as the standard
QCD contribution $\sigma_t$ from  $\qq, gg  \to \tt$. 
A precise measurement of the top cross section
therefore  provides an important check on the
model.  We shall quantify this in what
follows, using the same parameters as were  determined in Ref.~\cite{abfgm}
from a fit to the $R_{b,c}$ and large $E_T$ jet data.

The top cross section has been studied in the context of a variety
of new physics scenarios \cite{np}, especially since the original measurement
by the CDF collaboration gave a value somewhat higher than the
standard QCD prediction \cite{cdforig}. What distinguishes the present
study is that we are using a model whose parameters have already been
constrained, and therefore our predictions are on a firmer footing.

Our calculations are based on the $Z'$ model introduced in
Ref.~\cite{abfgm}, where full details can be found. Only a brief
summary is presented here. The neutral current sector of the
electroweak Lagrangian receives an additional
contribution
\begin{equation}
{\cal L}_{Z'} = {e \over 2 \sin\theta_W\cos\theta_W}
Z'^{\mu} \sum_f \bar\psi_f \gamma_{\mu}(\vp_f+ \ap_f \gamma_5)\psi_f
\end{equation}
where the vector and axial couplings are parametrized as
\begin{eqnarray}
\vp_u = x + y_u \; , && \ap_u = -x + y_u \nonumber \\
\vp_d = x + y_d \; , && \ap_d = -x + y_d \nonumber \\
\vp_l=\vp_{\nu} = 0\; ,&& \ap_l = \ap_{\nu} = 0\; ,
\end{eqnarray}
with $\vp_u = \vp_c = \vp_t$ etc.
Fixing the mass of the $Z'$ at $M_{Z'}=1\;\mbox{TeV}$,
the  parameters $x$, $y_u$ and $y_d$ 
can be adjusted to fit the measurements of $R_{b,c}$
while retaining the quality  of the Standard Model description of
other electroweak observables. The latter constraint forces the
leptonic couplings to be small, and following Ref.~\cite{abfgm}
we set them to zero. A fit to the LEP and SLC observables, including 
$R_b$ and $R_c$, yields~\cite{abfgm} a set of parameters 
which appears to be incompatible with the magnitude of the CDF
jet cross section. Moving these parameters within their errors, one is
able to obtain agreement with both the LEP/SLC observables and the CDF data
for the following values~\cite{abfgm}:
\begin{equation}
x=-1,\quad y_u = 2.2, \quad y_d = 0\; ,
\label{bestfit}
\end{equation}
which we will refer to as `final fit' in the following.
Note that a variation of $x$ between
$-1.5$ and $-0.5$ and of $y_u$ between $2$ and $4$  yields values for
observables which are still compatible with the experimental data~\cite{abfgm}.

\begin{figure}[htb]
\begin{center}
~ \epsfig{file=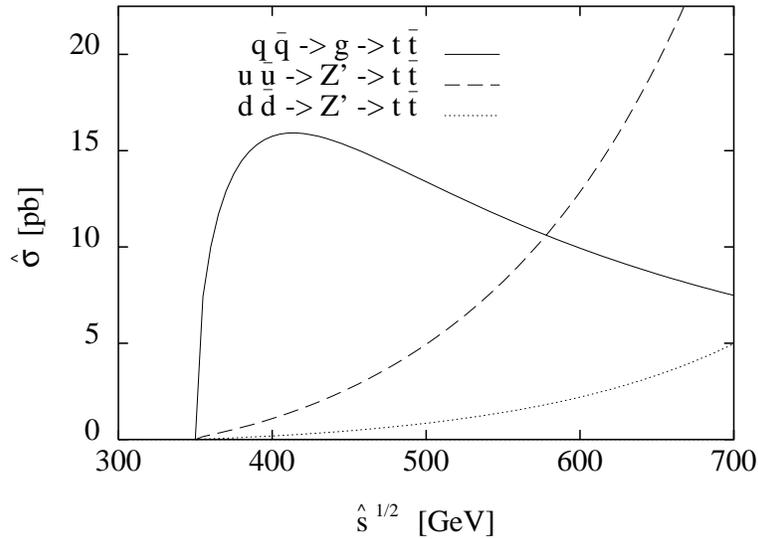,angle=-90,width=10cm}
\caption{Parton-level cross sections for the production of a
$t\bar{t}$ pair at leading order. 
The coupling parameters for the $Z'$ correspond to
the `final fit' of Ref.~\protect\cite{abfgm}.}
\end{center}
\end{figure}

The leading--order $\tt$ production subprocess cross sections from
standard QCD and from the anomalous $Z'$ contribution are:
\begin{eqnarray}
\hat\sigma_t(\qq\to\tt) & =& {4\pi\alpha_S^2\over 27 \shat}\;
\beta (3-\beta^2)\; , \nonumber \\
\hat\sigma'_t(\qq\to Z'\to \tt)  &=& {(G_FM_Z^2)^2\over 6\pi}\;
{\shat \over (\shat  - M_{Z'}^2)^2 +(\shat \Gamma_{Z'}/M_{Z'})^2 }
\nonumber \\
&& \times\ (\vp^2_q + \ap^2_q)\; \left[
{\beta\over 2}(3-\beta^2) \vp^2_t
+\beta^3\ap^2_t  \right] \; .
\label{eq:sub}
\end{eqnarray}
where $\beta^2 = 1 - 4m_t^2/\shat$ and the $Z'$ width is (for
$M_{Z'} \gg m_q$)
\begin{equation}
\Gamma_{Z'} = { G_F M_Z^2 \over 2\sqrt{2}\pi}\; 3 M_{Z'} \left[
\vp^2_u + \ap^2_u + \vp^2_d + \ap^2_d \right] \; .
\end{equation}
Figure~1 displays these parton-level cross sections as a function of
the subprocess centre-of-mass energy for $m_t=175 \;\mbox{GeV}$.
For the anomalous contribution, it is evident that 
only $u \bar u$ annihilation  will yield a sizeable
contribution to the  cross section. 

Our calculations of the corresponding $p\bar p$ cross sections use
the MRS(A$'$) parton distributions of Ref.~\cite{mrsap}, with
$\alpha_S(M_Z^2) = 0.112$. Note that approximately 90\% of the QCD
cross section comes from the $\qq\to\tt$ subprocess. We include
also the next-to-leading-order (NLO)
perturbative QCD corrections to (\ref{eq:sub}).
For the  standard QCD
$\qq,gg\to\tt$ cross sections these are taken from Ref.~\cite{nde}.
The NLO corrections to $\sigma'_t$ factor into two pieces:
\begin{equation}
\sigma'_{\rm LO+NLO} = \sigma'_{\rm LO} \otimes K_{DY} \otimes K_{Z'\to \tt}\; ,
\end{equation}
where $K_{DY}$ is the   $O(\alpha_S)$ Drell-Yan correction
\cite{dynlo} for $\qq\to Z'$ and $K_{Z'\to\tt}$ is the standard
$O(\alpha_S)$ correction for the decay of a $Z$-like vector boson
into heavy quarks \cite{heavynlo}.
Quantitatively, we observe that each of the  K-factors increases the
lowest-order 
cross section by about 15--20\%.
Using the `final fit' of (\ref{bestfit}), we find at
$\sqrt{s}=1.8\;\mbox{TeV}$ and $m_t=175\;\mbox{GeV}$:
\begin{eqnarray}
  \sigma'_{\rm LO} & =&  1.50 \;\mbox{pb}  \nonumber\\
  \sigma'_{\rm LO}\otimes K_{DY} & =&  1.74 \;\mbox{pb}
\nonumber\\ 
  \sigma'_t = \sigma'_{\rm LO}\otimes K_{DY}\otimes
K_{Z'\to \tt} & = & 
2.00 \;\mbox{pb},
\end{eqnarray}
to be compared to the Standard Model prediction of $\sigma_t=4.75 \;\mbox{pb}$.

\begin{figure}[htb]
\begin{center}
~ \epsfig{file=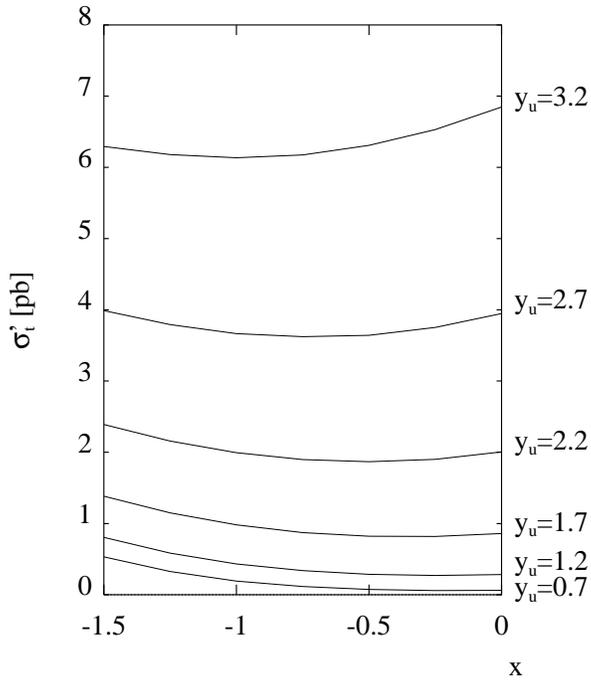,width=7.8cm}
\caption{Variation of $\sigma'_t$ with $x$, $y_u$.}
\end{center}
\end{figure}

The NLO $\sigma'_t$ cross section, for $\pp$ collisions at $\sqrt{s} =
1.8\; \TeV$ with $m_t = 175\; \GeV$ and MRS(A$'$) partons \cite{mrsap},
is shown as a function of the parameters
$x$ and $y_u$ in Fig.~2. The dependence on the third parameter $y_d$
is very weak. Note that  that $y_u<2$ is disfavoured
by the LEP/SLC data~\cite{abfgm}.  
The relative insensitivity to  the parameter $x$ evident
in the figure can be easily understood, as $x$ enters directly in the
$\uu, \dd \to Z'$ production
cross sections, see Eq.~(\ref{eq:sub}). An increase in
the production of $Z'$ bosons is however compensated by 
a larger amount of $Z'$ decays to $d$-type quarks. In contrast, an
increase in $y_u$ yields only an increase in $\uu\to Z'$, and
correspondingly in the overall top  cross section.

\begin{figure}[htb]
\begin{center}
~ \epsfig{file=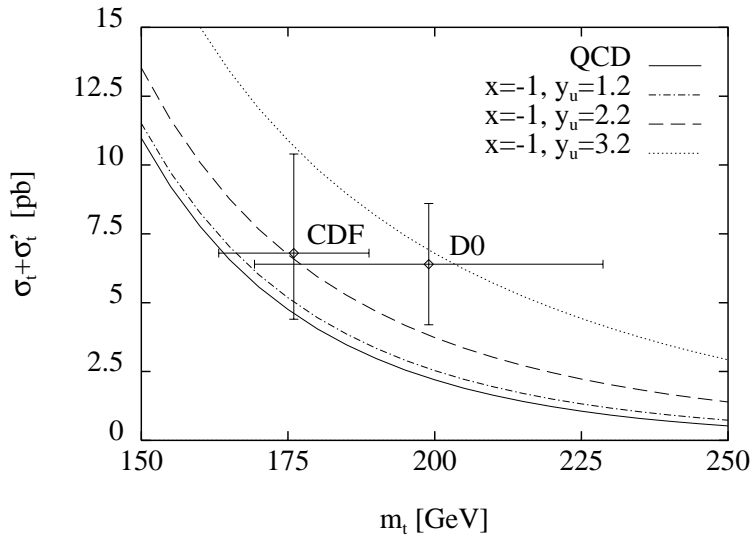,angle=-90,width=10cm}
\caption{Predictions for $\sigma_t+\sigma_{t'}$ as a function of $m_t$, with
data points from CDF and  D0.}
\end{center}
\end{figure}

Figure~3 shows the total cross section $\sigma_t+\sigma'_t$ as a function
of $m_t$ at the Tevatron collider. The soild line is the standard NLO
QCD prediction, the dashed line includes $Z'$-exchange with the
`final fit' (\ref{bestfit}) coupling parameters, and  the dot-dashed
(dotted)
line correponds to a smaller (larger) value for the  coupling parameter
$y_u$. The data points are from CDF \cite{cdf} and D0 \cite{dzero}.
Although the Standard Model prediction for $\sigma_t$ is still within
the range of the experimental errors, it can clearly be seen that the
additional $\sigma'_t$ contribution improves the agreement between
data and theory. Furthermore, more accurate measurements of $m_t$ and
the cross section, which should become available in the 
near future, will provide an additional constraint on the parameter $y_u$. 

The confirmation of an excess in the measured top cross section
must of course take 
into account the theoretical uncertainty in the Standard Model
prediction. There are three major sources of such uncertainty: unknown
higher-order perturbative corrections, the value of $\alpha_S$
 and parton distributions. A very complete study of this issue has recently
been performed in Ref.~\cite{cmnt} (see also earlier discussions in 
Refs.~\cite{wjsparis} and \cite{rkeglasgow}). The `best estimate'
 of the top  cross section (at $\sqrt{s} = 1.8\;\mbox{TeV}$)
 and its error from Ref.~\cite{cmnt} is
\begin{equation}
\label{eq:best}
\sigma_t = 4.75 {+0.63 \atop -0.68}\; \mbox{pb}\; .
\end{equation}
Note that the central value in (\ref{eq:best})
agrees with our result for $\sigma_t$ given above.
More generally, the error is approximately $\pm 15\%$ over the allowed
top mass range. The important point to note is that the `final fit' prediction
for $\sigma'_t$ is about three times larger than 
the error on the QCD prediction.

Given the uncertainties in the standard QCD prediction and in the data, 
it is 
important to investigate other properties of the final state
which could help distinguish an anomalous contribution to the
cross section. Examples include the angular distributions of the
top quarks and their decay products, as emphasized in Ref.~\cite{klane}.

\begin{figure}[htb]
\begin{center}
~ \epsfig{file=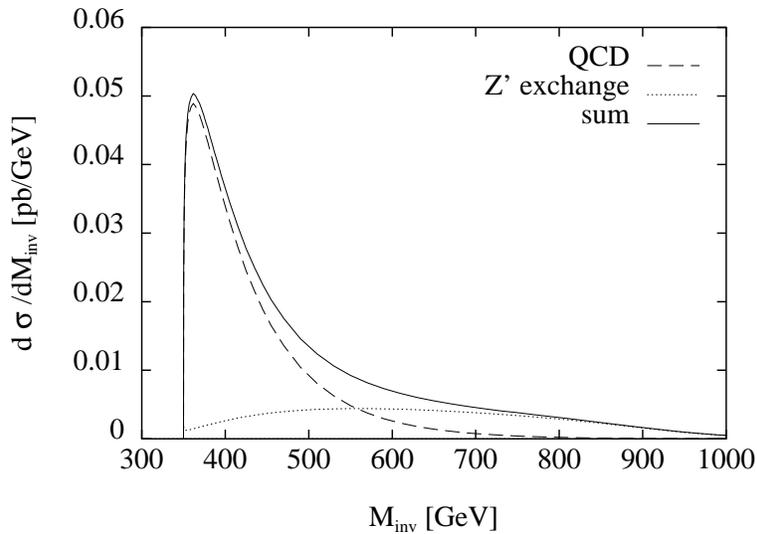,angle=-90,width=10cm}
\caption{Invariant mass distribution of $t\bar{t}$ final states at the
Tevatron.} 
\end{center}
\end{figure}

Notice in particular that the
rapidity distribution of the $t$ quark produced in
$\qq\to Z'\to \tt$ is {\it not} forward--backward symmetric, in contrast
to the standard production mechanism. However the simplest
discriminator of the anomalous and standard contributions is the
distribution in the invariant mass $M_{\rm inv}$
 of events containing $\tt$ pairs, shown in Fig.~4 for
 $\sqrt{s}=1.8\;\mbox{TeV}$, $m_t = 175 \; \GeV$ and the parameters of
(\ref{bestfit}). 
The dashed line denotes the Standard Model prediction, the additional
contribution due to the  exchange of the `final fit' $Z'$ is indicated by
the dotted line, and the solid line is the sum of these.
Just as for  the  excess in the single jet inclusive
distribution~\cite{cdfjet}, the $Z'$ contribution is visible as
an enhancement of the measured cross section at large invariant masses. 
Note that the final-state invariant mass at
next-to-leading order can include the contribution from
 additional gluon emission where appropriate. 
In practice, the invariant mass distribution of $\tt$ pairs will also depend
on kinematical cuts and the jet definition  used in the event
reconstruction. A detailed study of these effects
is  beyond the scope of this work.

In summary, we have shown that the new-physics model 
proposed in Ref.~\cite{abfgm}
to explain the anomalies in the measurements of $R_{b,c}$ at LEP/SLC
 and the CDF
large $E_T$ jet cross section  predicts
 a significant enhancement of the top quark
production cross section. The `final fit' estimate of the increase is 
about three times larger 
than the theoretical uncertainty in the standard prediction,
and should be readily observable given the expected increase 
in the precision of the experimental measurement in the future.

\section*{Acknowledgements}

\noindent  Financial support from the
Studienstiftung des deutschen Volkes (TG) is gratefully acknowledged.
This work was supported in part by the EU Programme
``Human Capital and Mobility'', Network ``Physics at High Energy
Colliders'', contract CHRX-CT93-0357 (DG 12 COMA).
\goodbreak


\begin{thebibliography}{10}

\bibitem{rbrc} The LEP collaborations ALPEH, DELPHI, L3, OPAL and the
LEP electroweak working group, preprint CERN-PPE/95-172 (1995).

\bibitem{cdfjet}
CDF collaboration: F.~Abe {\it et al.}, preprint FERMILAB-PUB-96/020-E
(1996).

\bibitem{abfgm}
G.~Altarelli, N.~Di~Bartolomeo, F.~Feruglio, R.~Gatto and M.L.~Mangano,
preprint CERN-TH-96-20 (1996).

\bibitem{clrv}
P.~Chiappetta, J.~Layssac, F.M.~Renard and  C.~Verzegnassi,
Marseille preprint CPT-96-P-3304 (1996).

\bibitem{np}
C.T.~Hill and S.J.~Parke, Phys. Rev. {\bf D49} (1994) 4454. \\
E.~Eichten and K.~Lane, Phys. Lett. {\bf B327} (1994) 129. \\
V.~Barger and R.J.N.~Phillips, Phys. Lett. {\bf B335} (1994) 510. \\
Jin~Min~Yang and Chong~Sheng~Li, Phys. Rev. {\bf D52} (1995) 1541. \\
K.~Lane, Phys. Rev. {\bf D52} (1995) 1546. \\
P.~Haberl, O.~Nachtmann and A. Wilch, Heidelberg preprint
HD-THEP-95-25 (1995).\\
K.~Whisnant, Bing-Lin Young and X.~Zhang, Phys. Rev. {\bf D52}
(1995) 3115. \\
D.~Atwood, A.~Kagan and T.G.~Rizzo, Phys. Rev. {\bf D52} (1995) 6264.

\bibitem{cdforig}
CDF collaboration:  F.~Abe {\it et al.}, Phys. Rev. Lett. {\bf 73}
(1994) 225; Phys. Rev. {\bf D50} (1994) 2966.

\bibitem{mrsap}
A.D.~Martin, R.G.~Roberts and
W.J.~Stirling, Phys. Lett. {\bf B354} (1995) 155.

\bibitem{nde}
P.~Nason, S.~Dawson and R.K.~Ellis, Nucl. Phys. {\bf B303} (1988) 607. \\
W.~Beenakker, H.~Kuijf, W.L.~van Neerven and J.~Smith,
Phys. Rev. {\bf D40} (1989) 54.

\bibitem{dynlo}
G.~Altarelli, R.K.~Ellis and G.~Martinelli, Nucl. Phys.
{\bf B143} (1978) 521; {\bf B146} (1978) 544(E);
{\bf B157} (1979) 461. \\
J.~Kubar-Andr\'{e} and F.E.~Paige, Phys. Rev.
{\bf D19} (1979) 221. \\
J.~Kubar, M.~le~Bellac, J.L.~Meunier and G.~Plaut, Nucl. Phys.
{\bf B175} (1980) 251.

\bibitem{heavynlo}
J.~Schwinger, {\it Particles, Sources and Fields}, Addison-Wesley, New York, 
1973.\\
J.~Jers\`{a}k, E.~Laermann and P.M.~Zerwas, Phys. Rev. {\bf D25}
(1982) 1218., {\bf D36} (1987) 310(E).\\
L.~Reinders, H.~Rubinstein and S.~Yazaki, Phys. Rep. {\bf 127} (1985) 1.

\bibitem{cdf}
CDF collaboration:  F.~Abe {\it et al.}, Phys. Rev. Lett. {\bf 74} 
(1995) 2626; Phys. Rev. {\bf D52} (1995) 2605.

\bibitem{dzero}
 D0 collaboration:  S.~Abachi {\it et al.}, Phys. Rev. Lett. {\bf 74} 
(1995) 2632; Phys. Rev. {\bf D52} (1995) 4877.

\bibitem{cmnt}
S.~Catani, M.L.~Mangano, P.~Nason and L.~Trentadue, preprint CERN-TH/96-21
(1996).

\bibitem{wjsparis}
W.J.~Stirling,
Proc.  Workshop on Deep Inelastic 
Scattering and QCD, Paris, April 1995, eds. J.-F.~Laporte and
Y.~Sirois, Ecole Polytechnique, Paris, p.91.

\bibitem{rkeglasgow}
R.K.~Ellis,
Proc. XXVII Int. Conf. on High Energy Physics, Glasgow, July 1994,
eds. P.J. Bussey and I.G. Knowles, Institute of Physics Publishing,
Bristol, Vol.II, p.1203. 

\bibitem{klane}
K.~Lane, Ref.~\cite{np}.

\end{thebibliography}
\end{document}